\documentclass[12pt]{article}
\usepackage[centertags]{amsmath}
\usepackage{amsfonts}
\usepackage{amssymb}
\usepackage{amsthm} 
\usepackage{newlfont}
\usepackage{epsfig}
\usepackage{amscd}

\newcommand{\beq}{\begin{equation}}
\newcommand{\eeq}{\end{equation}}
\newcommand{\ba}{\begin{array}}
\newcommand{\ea}{\end{array}}
\newcommand{\bea}{\begin{eqnarray}}
\newcommand{\eea}{\end{eqnarray}}

\begin{document}

\begin{center}
{\large \sc \bf A five-dimensional solitary-wave first order nonlinear PDE integrable by  dressing method.
}

\vskip 15pt

{\large  A. I. Zenchuk }

\vskip 8pt

\smallskip

{\it  Institute of Chemical Physics, RAS,
Acad. Semenov av., 1
Chernogolovka,
Moscow region
142432,
Russia}

\smallskip

\vskip 5pt


\vskip 5pt

{\today}

\end{center}

\begin{abstract}
We derive a five-dimensional nonlinear first order matrix PDE which is a generalization of the completely integrable 
(2+1)-dimensional $N$-wave equation. Similar to the $\bar\partial$-problem,  
our algorithm is based on the Fredholm-type linear integral equation with the kernel of  special form.

\end{abstract}

\section{Introduction}

The nonlinear Partial Differential Equations (PDE) integrable by the Inverse Spectral Transform Method
(ISTM-integrable PDEs) 
represent a large class of completely integrable nonlinear equations \cite{GGKM}.
The dressing method \cite{Sh}  as a special method for solving the nonlinear PDEs was  
developed in \cite{ZSh1,ZSh2} on the basis of Volterra type integral operators.
Later, another version of the dressing method  involving the Fredholm type integral operator ($\bar\partial$-problem)
was proposed in
\cite{ZM}.   The two families of ISTM-integrable PDEs are well  studied. These are the (1+1)- and (2+1)-soliton PDEs
(Korteveg-de Vries \cite{GGKM,KV}, Kadomtsev-Petviashvili \cite{KP}, Nonlinear Shreodinger \cite{ZSh3} 
and Devi-Stewartson \cite{DS}
equations should be 
mentioned as remarkable examples of such equations) and  the Self-dual type equations \cite{W,BZ}. 
Recently the equations associated with commuting vector 
fields were added to this list \cite{MS1,MS2}. Some  modifications of the $\bar\partial$-method 
increasing the dimensionality of (partially)
integrable PDEs are also known, see \cite{ZS_2007}. 

In this paper we consider a new modification of the dressing method based on the Fredholm-type operator and derive 
a   five-dimensional first order  nonlinear PDE with large solution space, 
although we do not show  its complete integrability. This  PDE has a limit to the classical 
ISTM-integrable   (2+1)-dimensional $N$-wave equation.

The structure of the paper is following. In Sec.\ref{Section:dressing}, we describe our 
modification of the dressing algorithm and represent  the general steps of the derivation of the nonlinear PDE.
The available solution space is discussed in Sec.\ref{Section:solution}. The main results are  formulated in 
Sec.\ref{Section:conclusion}.

\section{Dressing algorithm}
\label{Section:dressing}
We consider the $N\times N$ matrix function $R(x;\lambda,\mu)$ depending on 
two spectral parameters $\lambda$ and $\mu$ and
set of auxiliary parameters $x=(x_1,x_2,\dots)$, which will be 
the independent variables of the nonlinear PDEs.
We refer to the inverse of $R$  as the spectral function  $\Psi$:
\begin{eqnarray}\label{wave}
\Psi(x;\lambda,\mu)=R^{-1}(x,\lambda,\mu),
\end{eqnarray}
where 
\begin{eqnarray}
\int d\Omega(\nu) R(x,\lambda,\nu)R^{-1}(x;\nu,\mu) =I \delta(\lambda-\mu),
\end{eqnarray}
where $\Omega(\nu)$ is some scalar measure on the plane of complex spectral parameter $\nu$, $I$ is the  $N\times N$ 
identity operator.
The $x$-dependence can be introduced by the system of  linear PDEs
\begin{eqnarray}\label{QX}
R_{x_k}(x;\lambda,\mu)=r(x;\lambda) A^{(k)} q(x;\mu) + Q^{(k)}(\lambda,\mu),
\end{eqnarray}
where $F$ is a constant matrix, $Q^{(k)}$ are functions of spectral parameters independent on $x$.  
The latter assumption is taken for simplicity and, in principle, can be removed.
Notice that if $Q^{(k)}\equiv 0$, then we obtain (2+1)-dimensional ISTM-integrable 
$N$-wave equation  for the matrix field $v$,
\begin{eqnarray}\label{v}
v=\int d\Omega(\lambda) d\Omega(\nu)  q(x;\lambda) \Psi(x;\lambda,\mu) r(x;\mu).
\end{eqnarray}
The term $Q^{(k)}$ in (\ref{QX}) leads us to the higher-dimensional nonlinear PDEs.

The compatibility condition  of
system (\ref{QX}) yields 
\begin{eqnarray}
\Big(r(x;\lambda) A^{(k)} q(x;\mu)\Big)_{x_n} = 
\Big(r(x;\lambda) A^{(n)} q(x;\mu)\Big)_{x_k}
\end{eqnarray}
which can be  splitted into the following two equations:
\begin{eqnarray}
\label{qx}
&&
r_{x_n} A^{(k)}= r_{x_k} A^{(n)},\\\label{rx}
&&
A^{(k)} q_{x_n} = A^{(n)} q_{x_k}.
\end{eqnarray}

\subsection{Derivation of  spectral equation }
In this section we derive the spectral equation, i.e. 
PDE for a functions of spectral parameters $\Psi$. For this purpose, we  
differentiate  eq.(\ref{wave}) with respect to $x_k$ obtaining:
\begin{eqnarray}\label{V}
\Psi_{x_k}(x;\lambda,\mu)= - \chi(x,\lambda) A^{(k)} \tilde \chi(x;\mu)-
\int d\nu^2 d\tilde \nu^2 \Psi(x;\lambda,\nu)  Q^{k}(\nu,\tilde\nu) \Psi(x;\tilde \nu,\mu),
\end{eqnarray}
where
\begin{eqnarray}
 &&
 \chi(x,\lambda) = \int d\Omega(\nu)  \Psi(x;\lambda,\nu) r(x,\nu),\;\;\;
 \tilde \chi(x,\mu) =
\int d\Omega(\nu)  q(x;\nu)  \Psi(x;\nu,\mu).
\end{eqnarray}
For the sake of brevity,  we denote $*$  the integration over the ''inside pair'' of spectral parameters,
\begin{eqnarray}
f*g\equiv\int d\Omega(\nu) f(\lambda,\nu) g(\nu,\mu).
\end{eqnarray}

\subsubsection{Additional constraints for function $R$}

Having spectral equation (\ref{V})  we are not able to 
construct a complete system of nonlinear PDEs. 
Therefore we have to impose some additional relations on $R$ in the following form:
\begin{eqnarray}\label{BQ}
 Q^{(n)} * R-  R* Q^{(n)} = r C^{(n)} q ,
\end{eqnarray}
where 
$C^{(n)}$ are some diagonal constant matrices. 
Applying $\Psi*$ and $*\Psi$ to eq.(\ref{BQ}) we rewrite it as follows:
\begin{eqnarray}\label{red}
\Psi* Q^{(n)} -  Q^{(n)}* \Psi =\Psi* r C^{(n)} q* \Psi,
\end{eqnarray}

Compatibility of eqs.(\ref{BQ})
yields
\begin{eqnarray}\label{C1}
&&
Q^{(n)}* r C^{(m)}= Q^{(m)}* r C^{(n)},\\\label{C2}
&&
 C^{(m)} q* Q^{(n)} =C^{(n)} q* Q^{(m)},
 \\\label{commB}
&&
Q^{(n)} * Q^{(\tilde n)} = Q^{( \tilde n)} *Q^{(n)}.
 \end{eqnarray}
Compatibility of eqs.(\ref{BQ}) and (\ref{QX}) yields 
\begin{eqnarray}
 \label{C3}
&&
r_{x_k} C^{(n)}= Q^{(n)}* r A^{(k)} ,
\\\label{C4}
&&
C^{(n)}q_{x_k}= -A^{(k)} q*Q^{(n)}.
\end{eqnarray}
In virtue of eq.(\ref{BQ}), spectral equation (\ref{V}) reads 
\begin{eqnarray}\label{linV}
E^{(n)}:=\Psi_{x_n}  + \chi A^{(n)} \tilde \chi + \Psi^{(2)} * Q^{(n)} - \chi^{(2)} C^{(n)} \tilde \chi =0,
\end{eqnarray}
\begin{eqnarray}
\Psi^{(2)}=(\Psi F)*\Psi ,\;\;\; \chi^{(2)} = \Psi^{(2)}*r, \;\;\; \tilde \chi = q*\Psi.
\end{eqnarray}
System (\ref{linV}) together with constraint (\ref{red}) represent a spectral system for the nonlinear PDEs derived 
in the next subsection.

\subsection{Derivation of nonlinear PDEs}

First of all, deriving the nonlinear PDE  we  note 
that the term with $Q^{(n)}$  can be eliminated from eq.(\ref{linV}) 
owing to    condition (\ref{C1}). To use this condition, 
we apply $*r$ to eq.(\ref{linV}) and
consider 
the following combination: 
\begin{eqnarray}\label{VV}
&&
E^{(nm)}:= E^{(n)}*r C^{(m)} - E^{(m)}*r C^{(n)}= \\\nonumber
&&
\chi_{x_n} C^{(m)} -\chi_{x_m} C^{(n)}+
 \chi A^{(n)} v C^{(m)} - \chi A^{(m)} v C^{(n)}  - \chi^{(2)} C^{(n)} v C^{(m)}+\\\nonumber
&&
\chi^{(2)} C^{(m)} v C^{(n)} -\Psi*r_{x_n} C^{(m)} +\Psi*r_{x_m} C^{(n)} ,
\end{eqnarray}
where the field $v$ is defined in (\ref{v}).
Now, applying $q*$ to eq.(\ref{VV}) we obtain the nonlinear equation without spectral parameters:
\begin{eqnarray}\label{EE}
\hat E^{(nm)}:=\hat E^{(nm)}_0  - \hat F^{(nm)},
\end{eqnarray}
where
\begin{eqnarray}\label{E0}
&&
\hat E^{(nm)}_0   = v_{x_n} C^{(m)} -v_{x_m} C^{(n)} +
vA^{(n)} v C^{(m)} - v A^{(m)} v C^{(n)}  -\\\nonumber
&& v^{(2)} C^{(n)} v C^{(m)}+
v^{(2)} C^{(m)} v C^{(n)},\\\label{F}
&&
F^{(nm)}=\tilde \chi*r_{x_n} C^{(m)} -\tilde \chi*r_{x_m} C^{(n)}+ q_{x_n}*\chi C^{(m)} -
q_{x_m}* \chi C^{(n)} =\\\nonumber
&&
\tilde \chi*r_{x_1} A^{(n)} C^{(m)} -\tilde \chi*r_{x_1} A^{(m)} C^{(n)}+ 
A^{(n)} q_{x_1}*\chi C^{(m)} -
A^{(m)}q_{x_1}* \chi C^{(n)}.
\end{eqnarray}
Here the following  additional fields appear:
\begin{eqnarray}
\label{v2}
&&
v^{(2)}=q*\Psi*\Psi*r,\\\label{others}
&&
q_{x_1}*\chi , \;\;\;\tilde \chi*r_{x_1} 
\end{eqnarray}
and we assume $A^{(1)}=I$. Eq.(\ref{EE}) holds for any pair $(n,m)$, $n\neq m$. 
Putting $m=1$ in eqs.(\ref{EE}-\ref{F}) and assuming $C^{(1)}=1$ we write them in the form
\begin{eqnarray}\label{E02}
&&
E^{(n1)}_0   =v_{x_n}  -v_{x_1} C^{(n)} +
vA^{(n)} v  - v  v C^{(n)}  - v^{(2)} [C^{(n)}, v],\\\label{F2}
&&
F^{(n1)}=
\tilde \chi*r_{x_1} A^{(n)}  -\tilde\chi*r_{x_1}  C^{(n)}+ 
A^{(n)} q_{x_1}*\chi  -
q_{x_1}* \chi C^{(n)}.
\end{eqnarray}
The following combination of equations (\ref{EE}) is free of fields (\ref{others}):
\begin{eqnarray}\nonumber
&&
\left|
\begin{array}{ccc}
E^{(21)}_{\alpha\beta} & E^{(31)}_{\alpha\beta} & E^{(n1)}_{\alpha\beta}\cr
A^{(2)}_\beta -  C^{(2)}_\beta &A^{(3)}_\beta -  C^{(3)}_\beta & A^{(n)}_\beta -  C^{(n)}_\beta \cr
A^{(2)}_\alpha -  C^{(2)}_\beta &A^{(3)}_\alpha -  C^{(3)}_\beta & A^{(n)}_\alpha -  C^{(n)}_\beta 
\end{array}
\right| =0 \;\;\;\Leftrightarrow \\\label{nl}
&&
\left|
\begin{array}{ccc}
(E^{(21)}_0)_{\alpha\beta} & (E^{(31)}_0)_{\alpha\beta} & (E^{(n1)}_0)_{\alpha\beta}\cr
A^{(2)}_\beta -  C^{(2)}_\beta &A^{(3)}_\beta -  C^{(3)}_\beta & A^{(n)}_\beta -  C^{(n)}_\beta  \cr
A^{(2)}_\alpha -  C^{(2)}_\beta &A^{(3)}_\alpha -  C^{(3)}_\beta & A^{(n)}_\alpha -  C^{(n)}_\beta 
\end{array}
\right| =0,\;\;\alpha\neq\beta,\;\;n=4,5,\dots.
\end{eqnarray}
and 
\begin{eqnarray}\nonumber
&&
\left|
\begin{array}{cc}
E^{(21)}_{\alpha,\alpha} & E^{(n1)}_{\alpha\alpha}\cr
A^{(2)}_\alpha -  C^{(2)}_\alpha & A^{(n)}_\alpha -  C^{(n)}_\alpha 
\end{array}
\right| =0 \;\;\;\Leftrightarrow \\\label{nl2}
&&
\left|
\begin{array}{cc}
(E^{(21)}_0)_{\alpha\alpha}  & (E^{(n1)}_0)_{\alpha\alpha}\cr
A^{(2)}_\alpha -  C^{(2)}_\alpha & A^{(n)}_\alpha -  C^{(n)}_\alpha 
\end{array}
\right| =0,\;\;\;n=3,4,\dots.
\end{eqnarray}
The hierarchy (\ref{nl},\ref{nl2}) involves two matrix fields $v$ and $v^{(2)}$. 
In addition, eq.(\ref{nl}) is off-diagonal. Thus, the complete system of PDEs is represented, 
for instance by eq.(\ref{nl}) with $n=4,5$ and by eqs.(\ref{nl2}) with $n=3,4$. Therefore, the derived system of PDEs is 
five-dimensional.

 \subsubsection{Relations among  non-diagonal elements of $v$ and $v^{(2)}$ }
 Eq.(\ref{red}) generates additional constraint 
 for $\Psi^{(2)}$ and $\Psi$ as follows.
 Applying $*\Psi$ to eq.(\ref{red}) from the right we obtain
 \begin{eqnarray}
 \label{red2}
 \Psi*Q^{(n)}*\Psi - Q^{(n)} *\Psi^{(2)} = \chi C^{(n)} \tilde \chi^{(2)}.
 \end{eqnarray}
 Using eq.(\ref{red}) we can write it in the following form:
 \begin{eqnarray}\label{red3}
  \Psi^{(2)}*Q^{(n)} - Q^{(n)}*\Psi^{(2)} = \chi^{(2)} C^{(n)} \tilde \chi + \chi C^{(n)} \tilde \chi^{(2)}.
 \end{eqnarray}
 Now we apply $q*$ from the left and $*r$ from the right to obtain
 \begin{eqnarray}\label{red4}
\tilde E^{(n)}:= \tilde \chi^{(2)}*Q^{(n)}*r-q* Q^{(n)}*\chi^{(2)} - (v^{(2)} C^{(n)} v + v C^{(n)} v^{(2)})=0.
 \end{eqnarray}
 Owing to eqs.(\ref{C1}) and (\ref{C2}) the following combination of equations (\ref{red4}) is free of the 
 terms with $Q^{(n)}$:  
 \begin{eqnarray}
 \sum_{perm(k,n,m)} (C^{(k)} \tilde E^{(n)} C^{(m)} - C^{(k)} \tilde E^{(m)} C^{(n)}),
 \end{eqnarray}
 which yields the following algebraic relation between $v^{(2)}$ and $v$:
 \begin{eqnarray}\label{CCC}
 \sum_{perm(k,n,m)} \Big(C^{(k)} (v^{(2)} C^{(n)} v + v C^{(n)} v^{(2)}) C^{(m)} - 
 C^{(k)}  (v^{(2)} C^{(m)} v + v C^{(m)} v^{(2)}) C^{(n)}\Big)=0.
 \end{eqnarray}
 Since $C^{(1)}\equiv I$, we can rewrite eq.(\ref{CCC}) putting $m=1$ as follows:
 \begin{eqnarray}\label{CCC2}
 &&
 C^{(k)} v^{(2)} [C^{(n)},v] + C^{(k)} v [C^{(n)},v^{(2)}]-C^{(n)} v^{(2)} [C^{(k)},v] - C^{(n)} v [C^{(k)},v^{(2)}]+
 \\\nonumber
 &&
 [v^{(2)}C^{(k)},vC^{(n)}] +  [vC^{(k)},v^{(2)}C^{(n)}] =0 ,\;\;n\neq k.
 \end{eqnarray}

\section{Solutions of  system (\ref{nl},\ref{nl2}) with constraint (\ref{CCC2})}
\label{Section:solution}
In this section we find a particular form of  functions $Q$ associated with particular solutions of the system 
eq.(\ref{QX},\ref{BQ},\ref{qx},\ref{rx},\ref{C1}-\ref{C4}).
We take the following solution of  system (\ref{qx},\ref{rx}):
\begin{eqnarray}\label{sol1}
&&
q(x;\mu)= \int d\Omega(k) e^{-k \sum_i A^{(i)} x_i} q_0(k,\mu),\;\; 
r(x;\lambda)= \int d\Omega(k)r_0(\lambda,k)e^{k \sum_i A^{(i)} x_i},
\end{eqnarray}
where $d\Omega(k)$ is some measure on the complex plane of the parameter $k$.
Solution of eq.(\ref{QX}) compatible with constraint (\ref{BQ}) reads:
\begin{eqnarray}\label{R}
R(x;\lambda,\mu) = \int d\Omega(k) d\Omega(\tilde k) 
\frac{r_0(\lambda,k)e^{(k-\tilde k) \sum_i A^{(i)} x_i} q_0(\tilde k,\mu)}{k-\tilde k} +
\sum_i x_i Q^{(i)}(\lambda,\mu)+ 
I\delta(\lambda-\mu),
\end{eqnarray}
where $A^{(1)}$ is the identity matrix.

Now we consider solution of eqs.(\ref{C3})  and (\ref{C4}). First, assuming invertibility of $r_0$ and $q_0$ we obtain from 
eq.(\ref{C4}):
\begin{eqnarray}\label{Q}
 Q^{(n)}(\lambda,\mu) =   \int d\Omega(k) q_0^{-1}(\lambda,k) k C^{(n)} q_0(k,\mu).
\end{eqnarray}
Then eq.(\ref{C3})
 requires the following relation between $r_0$ and $q_0$:
 \begin{eqnarray}\label{r0}
 r_0(\lambda,k)*q_0(k,\mu)=I \delta(\lambda-\mu)
 \end{eqnarray}
 so that eqs.(\ref{C1}-\ref{commB}) become identities.

 \subsection{Degenerate functions $r_0(\lambda,k)$ and $q_0(k,\mu)$}
 Next, we consider the
 degenerate operators $r_0$ and $q_0$ with the purpose of construction the explicite solutions:
 \begin{eqnarray}\label{deg}
 r_0(\lambda,k) = \sum_{i} r_{1i}(\lambda)r_{2i}(k),\;\;\; q_0(k,\lambda) = \sum_{i} q_{1i}(k) q_{2i}(\lambda).
 \end{eqnarray}
 With $r_0$ and $q_0$ from (\ref{deg}),
eqs. (\ref{C3})  and (\ref{C4}) can not be solved in general and, consequently, the functions 
$v$ and $v^{(2)}$ can not be constructed. 
This fact defers our equations from the classical 
soliton equations, where the arbitrary  degenerate kernel of integral operator results in 
the explicite formulas for the solutions of PDEs. 
But (\ref{C3})  and (\ref{C4}) can be solved  if we take  scalar $r_{2i}$ and 
$q_{1i}$ in the form of $\delta$-functions:
\begin{eqnarray}\label{delta}
r_{2i}(k) = \delta(k-a_i),\;\;q_{1i}(k) = \delta(k-b_i),
\end{eqnarray}
where $a_i$, $b_i$ are scalar complex constants.
Then equations (\ref{C3}-\ref{commB}) require the following structure for $Q^{(n)}$:
\begin{eqnarray}
 Q^{(n)}(\lambda,\mu) =    \sum_i r_{1i}(\lambda) b_i C^{(n)}  q_{2i}(\mu),\;\;\;
\end{eqnarray}
with
\begin{eqnarray}
 q_{2i}*r_{1j} =I\delta_{ij}.
\end{eqnarray}

\subsubsection{Explicite expression for  field $v$}
For the degenerate functions (\ref{deg},\ref{delta}), we can write the kernel $R$ (\ref{R}) in the form:
\begin{eqnarray}
R(\lambda,\mu) = \sum_{i,j} r_{1i}(\lambda) R^{(0)}_{ij} q_{2j}(\mu)+I\delta(\lambda-\mu),
\\\nonumber
R^{(0)}_{ij}= \frac{e^{(a_i-b_i)\sum_m A^{(m)} x_m}}{a_i-b_j} + \sum_n x_n C^{(n)} b_j \delta_{ij},
\end{eqnarray}
where $\delta_{ij}$ is the Kronecker symbol.
Substituting this $R$ into the evident equation  $(R*\Psi)(\lambda,\mu) = I\delta(\lambda-\mu)$ we obtain
\begin{eqnarray}\label{Psideg}
 \Psi(x;\lambda,\mu) =
 I\delta(\lambda-\mu) - \sum_{i,j} r_{1i}(\lambda) R^{(0)}_{ij}(x) \Psi_{j}(x;\mu)  
\end{eqnarray}
where $\Psi_j = q_{2j}*\Psi$.
Applying $q_{2k}*$ to eq.(\ref{Psideg})  from the left we obtain the linear algebraic  equation for $\Psi_j$:
\begin{eqnarray}\label{Psideg2}
 \Psi_k(x;\mu) =
q_{2k}(\mu)   - \sum_{j}  R^{(0)}_{kj}(x) \Psi_{j}(x;\mu),\;\;k=1,2,\dots,
\end{eqnarray}
or
\begin{eqnarray}\label{Psideg3}
\sum_j  \hat R_{kj} \Psi_j(x;\mu) =
q_{2k}(\mu)  ,\;\;\;\hat R_{kj}= I\delta_{kj} + R^{(0)}_{kj}.
\end{eqnarray}
which yields 
\begin{eqnarray}\label{Psij}
\Psi_j(x;\mu) = \sum_k (\hat R^{-1}(x))_{jk} q_{2k}(\mu),\;\;\;
\end{eqnarray}
where $\hat R$ has the following block structure: $\hat R =\{\hat R^{(0)}_{kj}\}$.
Substituting (\ref{Psij}) into eq.(\ref{Psideg}) we obtain
\begin{eqnarray}\label{Psideg4}
 \Psi(x;\lambda,\mu) =
 I\delta(\lambda-\mu) - \sum_{i,j,n} r_{1i}(\lambda) R^{(0)}_{ij}(x)  
 (\hat R^{-1}(x))_{jn} q_{2n}(\mu). 
\end{eqnarray}
Now, applying $q*$ and $*r$ to (\ref{Psideg2})  we obtain expression for $v$:
\begin{eqnarray}\label{vsol}
v=\sum_i e^{(a_i-b_i)\sum_m A^{(m)} x_m}  + \sum_{i,j,n}  e^{-b_i \sum_m A^{(m)} x_m}  
R^{(0)}_{ij}(x)  
 (\hat R^{-1}(x))_{jn} e^{a_n \sum_l A^{(l)} x_l} .
\end{eqnarray}
Solution of  form (\ref{vsol}) can be  referred to the solitary-wave solutions.
Expression for $v^{(2)}$ can be obtain in a similar way using its definition (\ref{v2}). We do not represent it here.

 \section{Conclusion}
 \label{Section:conclusion}
 
 In this paper we propose a modification of the dressing method based on the Fredholm-type 
 integral equation with the kernel of special type which, along with the Cauchy term exponentially depending on $x$, 
 involves a term linear in $x$. This term  allows us to increase the dimensionality 
 of the associated nonlinear equations. 
 We derive the five-order system of the first order PDEs (\ref{nl},\ref{nl2})  which is a multidimensional 
 extension of the classical (2+1)-dimensional 
 ISTM-integrable $N$-wave equation. This equation is supplemented by the algebraic constraint (\ref{CCC2}) 
 We show that these equation posses a rich family of solutions including the explicite solutions which can be produced by the degenerate 
 kernel of  integral operator.
 Further increase of dimensionality is possible and will be discussed later.
 
 The work is partially supported by  the Program for Support of Leading Scientific Schools (grant No. 3753.2014.2), 
 and by the RFBR (grant No. 14-01-00389).


\end{document}